\documentstyle[12pt,aasms4]{article}

\newcommand{\be}{\begin{equation}}
\newcommand{\ee}{\end{equation}}
\newcommand{\etal}{{\it et al.}}
\newcommand{\hmp}{h^{-1}Mpc}
\newcommand{\bef}{\begin{figure}}
\newcommand{\eef}{\end{figure}}
\def\spose#1{\hbox to 0pt{#1\hss}}
\def\ltapprox{\mathrel{\spose{\lower 3pt\hbox{$\mathchar"218$}}
 \raise 2.0pt\hbox{$\mathchar"13C$}}}
\def\gtapprox{\mathrel{\spose{\lower 3pt\hbox{$\mathchar"218$}}
 \raise 2.0pt\hbox{$\mathchar"13E$}}}
\def\inapprox{\mathrel{\spose{\lower 3pt\hbox{$\mathchar"218$}}
 \raise 2.0pt\hbox{$\mathchar"232$}}}

\slugcomment{Subimitted to Astrophys. J. Letters}

\lefthead{Joyce, Montuori and Sylos Labini}
\righthead{Fractal correlations in the CfA2-South redshift survey}

\begin{document}

\title{ Fractal correlations in the CfA2-South redshift survey }

\author{ 
M. Joyce \altaffilmark{1,2,3} 
M. Montuori\altaffilmark{1}  and 
F. Sylos Labini \altaffilmark{1,4}
}
\altaffiltext{1}{INFM Sezione Roma 1, Universit\'a di Roma 
"La Sapienza" P.le A. Moro 2, 00185 Roma, Italy }
\altaffiltext{2}{INFN Sezione Roma 1, Universit\'a di Roma 
"La Sapienza" P.le A. Moro 2, 00185 Roma, Italy}
\altaffiltext{3}{School of Mathematics, 
Trinity College, Dublin 2, Ireland.}
\altaffiltext{4}{ D\'ept.~de Physique Th\'eorique, Universit\'e de Gen\`eve, 
24, Quai E. Ansermet, CH-1211 Gen\`eve, Switzerland}

\begin{abstract}
We report our analysis of the properties of galaxy 
clustering for a new redshift sample of galaxies, the 
CfA2-South catalog, using statistical methods which do
not rely on the assumption of homogeneity.
We find that, up to $\sim  20 \hmp$, which is the largest scale 
to which correlation properties can be reliably inferred, the galaxy 
clustering is scale-invariant and characterized by a fractal 
dimension $D=1.9 \pm 0.1$. Further there is no statistical evidence 
for homogeneity at any of the larger scales (up to $\sim 150 \hmp$) 
probed more weakly by the catalog. These results means that characteristic 
``correlation lengths'' for the clustering of galaxies derived using 
standards methods of analysis are not meaningful.  
Further the results are very consistent with those obtained
from many other catalogs using the methods adopted here, which 
show the $D \approx 2$ fractal continuing to beyond 
$100 \hmp$. The incompleteness of the relevant data 
conjectured by various authors to give rise to such behaviour
is therefore proved to have no significant effect 
(up to $20 \hmp$) on the measured correlations.
\end{abstract}


\section{Introduction}
That the galaxy distribution exhibits fractal properties has been
well established for almost twenty years (\cite{man77,pee80}).
Controversy still surrounds questions concerning
the range of the fractal regime, the value of the fractal dimension 
and the eventual presence of a 
cross-over to homogeneity (\cite{pmsl97,dav97,wu98} and references therein).
In this letter we present the answers to these questions provided
by a statistical analysis of a new large redshift sample of 
galaxies, the CfA2-South survey (\cite{huchra98}).

Red-shift catalogs such like this one have been, and continue
to be,  extensively analyzed with the standard 
two-point correlation function $\xi(r)$ method (\cite{pee80})
which involves normalizing to a mean density extracted
from the catalog. Indeed the predictions of all standard type
theories of structure formation are usually framed in terms of 
these statistics (or related ones such as the power spectrum).
One result of this analysis is that up to a small scale 
($\sim 10 \div 20 \hmp$) power-law type behaviour is observed 
in $\xi(r)$ which is interpreted to indicate fractal behaviour
with a characteristic fractal dimension $D \approx 1.3$.  
The central result however is the derivation of  
a characteristic length, the ``correlation length'',
$r_0 \approx 5 \hmp$, which should mark the end of a fractal
behaviour and the onset of a trend towards homogenization: 
It is defined as the scale at which the
the density fluctuations become of the same order as the 
average density, and at a few times this distance the 
fluctuations have small amplitude i.e the density tends to 
some well-defined asymptotic homogeneous value. 
This trend to homogeneity at such a scale is in apparent 
agreement with the structure-less angular data, but it is puzzling  
with respect to the much larger structures observed in the 3-d data. 
From this perspective it seems that the presence or absence  
of structures in the data is irrelevant for the determination of 
$r_0$.  
 
Puzzled by this apparent anomaly Pietronero and collaborators 
(\cite{pie87,cp92,slmp98}) reconsidered the analysis of galaxy 
correlations with such methods, adopting instead a procedure
in which the correlations are characterized with statistics which
are appropriate whether there is homogeneity or not, and which 
allow the underlying assumption of homogeneity in the standard
analysis to be tested. The conclusion of such considerations
applied to various galaxy surveys such as CfA1, SSRS1, 
IRAS, APM-Stromlo, Perseus-Pisces and LEDA (\cite{slmp98}) 
is that there is in fact no statistical evidence for the 
assumed homogeneity on the scales probed by any of these catalogs
(up to $\sim 100 \hmp$ for the LEDA catalog), and 
that the derived ``correlation length'' therefore has no 
meaning except in relation to the particular sample.  
On the other hand there is indeed 
clear fractal behaviour at the scales probed by all these
catalogs, but it is correctly characterized by a fractal
dimension $D\approx 2$ rather than the smaller value derived
in the standard analysis. These results have been questioned 
on the basis of the statistical validity of the 
catalogs (\cite{dav97}). Various authors have in fact proposed 
that the incompleteness of these data may lead to an apparent 
fractal behaviour. In this context the catalog we analyse in this
paper is important as it is a new and very accurate one, which does
not suffer from the possible problems indicated with
other surveys. Indeed it has been analysed extensively using the
standard methods (\cite{par94}) and is frequently used as a constraint
on theories of structure formation.

Rather than being tested directly with methods such as those
used here, the assumption of homogeneity is often 
defended with the assertion that the ``correlation length''  
is indeed a real physical scale characterizing the clustering 
of galaxies,  because  it shows the stability in different
samples indicative of homogeneity and does not show the
behaviour which would be associated with a fractal
distribution. In practice, however, correlation lengths 
are not observed to be very stable. For example in the
SSRS2 catalog Benoist et al. (1996) have measured 
different values of $r_0$ in the range $[4,15] \hmp$.
In the CfA catalog as analysed by \cite{par94}
there is also a large variation of the measured 
correlation length, which we will discuss below.
It is here that concept of ``luminosity selection bias'' 
enters (\cite{dav88,par94,ben96}): 
Galaxies  of different brightness are supposed to be clustered 
differently, and it is this which is proposed as
the physical explanation of the observed variation,
rather than the absence of real underlying homogeneity. 
Rather than there being one real scale characterizing 
the correlations of galaxies, there is then an
undetermined number of such scales. 
In our discussion below we will pay particular attention
to this point, showing that the variation of the ``correlation
length'' in the analysis of Park et al. is perfectly consistent
with our results, and that the account of luminosity selection
bias invoked to explain it is therefore unnecessary.



We first describe our methods of analysis before proceeding to
the catalog and our results. Essentially we make use of
a very simple statistic (\cite{pie87,cp92,slmp98}) which
is appropriate for characterizing 
the properties of regular as well as irregular distributions.
This is the conditional
average density, defined as
\be 
\label{e4} 
\Gamma(r) = 
\left \langle \frac{1}{S(r)} \frac{dN(<r)}{dr} \right \rangle 
\ee 
where $dN(<r)$ is the number of points in a shell 
of radius $dr$ at distance $r$ from an occupied point and 
$S(r)dr$ is the volume of the shell. The average indicated
by the angle brackets is over all the occupied points contained 
in the sample. Another related quantity we use is 
\be
\label{eq1b}
\Gamma^*(r)
 = \frac{3}{4 \pi r^3} \int_{0}^{r} \Gamma(\rho) d^3 \rho \;,
\ee
which is just the conditional density in 
a sphere of radius $r$ rather than in a spherical shell 
as for $\Gamma(r)$ \footnote{Note that in real samples
the integral in eq.\ref{eq1b} has a lower cut-off
different from zero.}. It is thus a more stable
global quantity which smoothes out the rapidly varying 
fluctuations which may appear in $\Gamma(r)$.

These statistics are suitable both for the study of the 
approach (if any) to homogeneity of a given distribution,
as well as for the identification of fractal properties.
The former is indicated by the tendency to a constant 
as a function of distance $r$, while the latter will be 
indicated by a simple power-law behaviour. Specifically for a 
scale invariant distribution of points with fractal dimension
$D$ we have:  
\be 
\label{ee5} 
\Gamma(r) = \frac{BD}{4\pi } r^{-\gamma} \;  \quad {\rm and} \quad 
\Gamma^*(r)= \frac{3B}{4\pi} r^{-\gamma}
\ee 
where $\gamma=3-D$ and $B$ is the lower cut-off, characterizing
the end of self-similarity in the finite set. ($B$ is the 
number of galaxies contained in a ball of radius $1 \hmp$ around 
a given  galaxy.)

In the analysis of the correlation properties
of a real system with these  statistics 
there are two important physical scales defining
the range in which one can infer real correlation
information from them:

(i) The upper cut-off $R_s$ up to which
the statistic can be calculated. It is simply
the size of the largest sphere around any galaxy 
which can be inscribed inside the sample volume, 
since the conditional density is computed 
in complete shells (\cite{cp92,slmp98}).
It depends on the survey geometry - 
on the solid angle of the survey and the effective depth 
of the particular sub-sample we analyse. As one approaches 
this scale both the number of points being averaged over 
and the separation of these points drops, so that one is
no longer computing a real average correlation.
We will return to this point below in discussing our
results.

(ii) A lower cut-off $\langle \Lambda \rangle$,
which is related to the number of points contained 
in the sample. It is simply the scale below which
the behavior of the conditional density is 
dominated by the sparseness of the points. In 
$\Gamma(r)$ this regime is characterized by highly
fluctuating behaviour, while in the $\Gamma^*(r)$ 
it is manifest as $1/r^3$ decay away from any finite 
value. 

 
The problems of the standard analysis can readily be seen 
from the fact that, for the case of a fractal distribution,
the standard ``correlation function'' $\xi(r)$ and
``correlation length '' $r_o$ are just  
\be 
\label{e7} 
\xi(r) = \frac{\Gamma(r)}{<n>} -1 \quad {\rm and} \quad 
r_o = \left( \frac{BD}{8 \pi <n>} \right) ^{\frac{1}{\gamma}} 
\ee 
where $<n>$ is the average density in the sample as estimated
in the calculation of $\xi(r)$.  In a fractal $<n>$, and therefore 
$r_0$,  are quantities which characterize not intrinsic properties
of the distribution, but the particular sample one is considering. 
For example, in a 
sample of depth $R_s$
the density is, on average,  $<n>= \Gamma^*(R_s)$. Thus
for a $D=2$ fractal, the {\it average} value in
such samples is $r_0 = R_s/3$. 
Further, for the case of a fractal, $\:\xi(r)$ is a power law only for
$r \ll r_0$. 
The deviation from this behaviour at large scales is 
again just due to the size of the observational sample 
and does not correspond to any real change
of the correlation properties. 
The log derivative of eq.\ref{e7} with respect to $\log(r)$ is
\be
\label{xi4}
\gamma'=\frac{d(\log(\xi(r))}{d\log(r)}
 = - \frac{2 \gamma r_0^{\gamma} r^{-\gamma}} 
{2 r_0^{\gamma} r^{-\gamma} -1}
\ee
where $r_0$ is defined by $\xi(r_0) = 1$.
The tangent
to $\xi(r)$ at $r=r_0$ has a slope $\gamma'=-2\gamma$.
It is thus clear that, even if the
distribution has fractal properties, it is very difficult
to recover the correct slope from the study of 
the $\xi(r)$ function.

  
We now turn to the data.
The CfA2 South galaxy sample (\cite{huchra98})
contains 4390 galaxies with magnitude
$m_{B(0)} \le 15.5$ covering
$20^h \le \alpha \le 4^h$ in right
ascension and 
$-2.5^{\circ} \le \delta \le 90^{\circ}$ 
in declination. This part of the sky includes regions where 
the galactic extinction has been measured to be important,
which needs to be corrected for. For the purposes of comparison 
of our results with the analysis carried out by \cite{par94} 
we perform these corrections in exactly the same way as
these authors, by excluding the same 
regions: $20^h \le \alpha \le 21^h$ ;
$3^h \le \alpha \le 4^h$  ; $21^h \le \alpha \le 2^h$  
and $b >-25^o$ and  $2^h \le \alpha \le 3^h$  
and $b >-45^o$ where $b$ is the galactic latitude.
The catalog contains a large supercluster
called {\it the Perseus-Pisces chain}
and it reveals large voids in the foreground and 
background of this supercluster. The complex galaxy structures
found in this catalog have been described by \cite{huchra98}.
Note that all the wide angle redshift surveys 
(i.e. CfA1, SSRS1, SSRS2, Perseus-Pisces) have shown 
comparable fluctuations, which indeed represent the
the subject of our analysis.

A redshift survey limited in this way by apparent magnitude 
has a systematic selection effect: At large distances only 
the brighter galaxies  are included in the sample, while at 
small  distances there are also the fainter ones. To
avoid this selection bias it is a standard procedure to 
consider the so-called volume-limited (VL) samples 
(e.g. \cite{dp83}), in which one chooses an upper cut-off
in distance and takes only all the galaxies which are 
bright enough to be seen up to this distance. 
In Tab.\ref{tabvl} we report the characteristics of the 
VL samples we have considered, characterized by  successive 
limits in absolute magnitude with a step of $0.5$ magn. 
The distances have been computed 
from the linear Hubble law ($r = cz/H_0$ and 
$H_0=100h km/sec/Mpc$ where $0.5<h<1$) and the absolute 
magnitude from the standard relation 
$M = m - 5 \log_{10} r(1+z) -25 - Kz$, where for the 
$K$ correction we have taken $K=3$ (as in \cite{par94}).
The results of our analysis are extremely weakly 
dependent on this latter correction, as they are on
modifications to the $r-z$ relation corresponding to
different cosmological models, simply because the 
maximum red-shift is so small ($z < 0.05$).

 
In  fig.\ref{fig1} the conditional average density $\Gamma^*(r)$ is 
plotted for each of these five samples.
The error bars displayed correspond to the variance 
on 20 bootstrap resamplings of each sample\footnote{For a sample 
with $N$ points, a random
integer is generated between $1$ and $N$ for
each point. Those points with the same number
are discarded to produce the sub-sample.}.
The behaviour of these errors can also be 
understood in terms of the two limits we discussed above:
At small distances they are large because the results for the 
average quantities are dominated by a few points
which make the average density non-zero.
In some cases they are large at $r \sim R_s$ 
because here one is averaging about just a few
well separated galaxies, one of which is removed 
in the resampling. It is important to note however 
that although the behaviour of these errors is related to  
the two cut-offs, they {\it cannot} be taken to
be measurement errors for these effects, which are
systematic. At small scales the real behaviour of
the conditional density is determined by the real 
lower cut-off and is intrinsically highly fluctuating
no matter how well sampled; at large scales no resampling 
of the points around a single galaxy can tell us what
the intrinsic variance is in the quantity
measured from different independent points. 

An examination of the figure shows that beyond
a scale, which grows with the depth of the VL sample,
there is in each case a well defined power-law, until 
a scale near to the upper cut-off $R_s$ at which, 
in some samples, it shows a deviation towards a 
flatter behaviour. The first scale is just the
lower cut-off $\langle \Lambda \rangle$ due to
sparseness discussed above. It can be checked 
quantitively that its increase in the deeper samples
scales with the growing mean distance between points.
To perform a fit to these curves we also need to
take account of the systematic effect as one
approaches $r \sim R_s$, not included in the bootstrap
errors, due to the non-averaging. In principle we cannot
know how large the error at this scale is since we do not
know the real variance in the density at this scale.
The criterion we use here to place an upper cut-off
up to which we assume this systematic effect is not 
important is  a simple one: We require that the average 
distance between the centres of the spheres we are averaging 
over at the relevant depth $r$ be such that the 
spheres do not overlap. The quantitative meaning of
this can be read off from the figure inserted in figure one,
as the point where the average distance becomes equal
to twice the depth. Obviously this scale $R_u$ grows with
sample size and we see it reaches a maximum of about
$20 \hmp$ in our deepest VL sample. 

\placefigure{fig1}


In each of the sample we perform a best fit 
to the dimension $D$ in the range of scales 
$[\langle \Lambda \rangle,R_u]$ (see Tab.1).
Our result is that the dimension is $D = 1.9 \pm 0.1$ 
in this range of scales probed by the samples 
i.e. from $0.5 \hmp$ to $20 \hmp$. 

The normalization of the conditional average density
in different VL samples depends on the luminosity
selection function of the sample considered.
The procedure to perform such a normalization has been
described in \cite{slmp98}. 
In short we assume that the  joint luminosity  
and space   density can be written as
\be
\label{norm1}
\nu(M,r) = \Gamma(r) \cdot \phi(M) = 
\frac{DB}{4\pi}r^{D-3} \cdot \phi(M)
\ee
where the luminosity function $\phi(M)$ 
has been normalized to unity 
($\int_{-\infty}^{M_{min}} \phi(M) dM =1$
and $M_{lim}$ is the faintest absolute magnitude contained
in the sample).
We can associate to a VL sample limited at $M_{VL}$ 
a luminosity factor:
\be
\label{norm2}
\Phi(M_{VL}) = \int_{-\infty}^{M_{VL}} \phi(M)dM \;.
\ee
The normalization of the space density is then
\be
\label{norm3}
\Gamma_{norm}(r) =\frac{ \Gamma(r)}{\Phi(M_{VL})} =
\frac{DB}{4\pi}r^{D-3} \cdot  \frac{1}{\Phi(M_{VL})}
\ee
From eq.\ref{norm3}, we can estimate the parameter $B$ of
the distribution. We find
$B \approx (12 \pm 2  \hmp)^{-D} \; ,$
which agrees very well with the value found
in various other catalogs (\cite{slmp98}). This
concordance clearly shows that the results derived
in these catalogs (which are in good agreement with 
each other) are not significantly effected by 
incompleteness or other imperfections, at least at the
scales probed by the present analysis. The dismissal
on these grounds of such very robust and self-consistent
results thus seems (even at larger scales) rather
inappropriate.

Beyond $\sim 20 \hmp$ we cannot reliably infer average
properties of the galaxy distribution from the CfA2 
South catalog alone. We can see however
that, at least to $\sim 30 \hmp$, there is evidence of
a clearly fluctuating behaviour consistent with the
continuation of the fractal. We can further examine 
unaveraged quantities like the number counts $N(<r)$
from the origin. These show the highly fluctuating
behaviour typical of a fractal, with slopes ranging
from $3.5$ to $2$ (corresponding to a global overdensity
on the scale of the survey).  Are we potentially 
missing evidence for a real
flattening in the conditional density by assuming that
the variance is large i.e could the deviations at 
$r \sim R_s$ from the simple power-law in a few samples 
be real averaged ones?
Could the slopes closer to $3$ in a few samples in the 
number counts be real average behaviour? One simple way in
which one can discount this possibility is to show that
the particular behaviour is not an average one, even for
the sample, by showing it to be associated with a particular 
feature within the sample. In the present case the evidence
for this is very clear-cut: These behaviours are all
associated with the main overdensity in the sample, the
Perseus-Pisces supercluster. The judgement one might 
make at first glance, that this catalog is shows no tendency
to homogeneity simply because of this feature, is quite 
simply correct.

Finally we return to the standard $\xi(r)$ analysis,
in particular as applied by \cite{par94} to the CfA2
catalog. Our findings
of power-law correlations without evidence for a cut-off
imply that  the normalization to a 
mean density to derive a ``correlation length'' is 
{\it conceptually} flawed. Calculationally,
however, there is nothing wrong with deriving such 
a scale, and the results should
be perfectly consistent (numerically) with those given here. 
In Table 1 we list the simple volume density 
$\langle n \rangle$ in each of the VL samples.
Using this as our normalizing density in equation (\ref{e7}),
and the values of $B$ and $D$ from the measured $\Gamma^*(r)$,
we obtain $r_o$ in the range $5 \div 10 \hmp$.  
These values are very compatible with those which can be 
read off from Figure 10 in Park et al.(for the whole CfA survey)
which gives values for $r_o$ in the VL samples shown 
of approximately $7 \hmp$ and $9 \hmp$. As we discussed above
the {\it average} $r_o$ predicted in a continuing 
$D=2$ fractal for such samples is $r_o=R_{VL}/3$.
The ratio of the measured $r_o$ therefore agrees 
strikingly well with this average scaling since the two samples 
have respective depths $101 \hmp$ and $130 \hmp$. 
The difference between the absolute values 
for $r_o$ and the {\it average} $r_o$ implies 
that the CfA2 survey would correspond in this case
to an overdensity by a factor of four (relative to the
average  density $\Gamma^*(R_{VL})$ at the depth of 
the survey). To 
invoke (as \cite{par94} do)
the additional hypothesis of luminosity selection
bias to explain all the variation of the observed $r_o$
is in our view very problematic unless
one has first clarified the role of the intrinsic variance 
in the densities to which one is normalizing to obtain $r_o$.

We conclude by noting again that our results on this survey
permit only the establishment of a fractal with $D \approx 2$ 
to $\sim 20 \hmp$. While we find no statistical evidence in this
survey for the hypothesis of homogeneity up to the largest
scales probed ($\sim 150 \hmp$), we cannot exclude that the
fluctuations we see in non-averaged quantities rule out any
trend to homogeneity at such scales. The question of the true
properties of the galaxy distribution at such scales will only
be definitively clarified with the advent of the much 
larger forthcoming red-shift surveys (2dF and SLOAN). 
The application of methods such as those used here, which do 
not assume homogeneity, is essential for the resolution of
this fundamental observational question.


\acknowledgments
We warmly thank L. Pietronero  for useful discussions 
and collaborations.
This work has been partially supported by the 
EEC TMR Network  ``Fractal structures and  
self-organization''  
\mbox{ERBFMRXCT980183} and by the Swiss NSF.

\clearpage
\begin{table}
\begin{tabular}{lllllllll} 
\hline
Sample & $R_{VL}$  & $M_{VL}$  & N 
& $R_s$ & $\langle n \rangle$ &  $\langle \Lambda \rangle $ & $R_u$ & D \\
\hline 
VL185 & 60.2  & -18.5 & 724  & 13.0  & $1.3 \cdot 10^{-2}$ & 0.5 & 7  & 2.0 \\
VL19  & 74.9  & -19.0 & 622  & 18.0  & $6.0 \cdot 10^{-3}$ & 0.7 & 8  & 1.8 \\
VL195 & 92.9  & -19.5 & 520  & 18.5  & $2.5 \cdot 10^{-3}$ & 1.2 & 12 & 1.9 \\
VL20  & 115   & -20.0 & 292  & 24.0  & $7.5 \cdot 10^{-4}$ & 3.0 & 15 & 1.9 \\
VL205 & 141.7 & -20.5 & 132  & 29.6  & $2.0 \cdot 10^{-4}$ & 6.0 & 20 & 1.8 \\  
\hline
\hline
\end{tabular} 
\caption{Volume limited samples
of the   CfA2-South. 
$R_{VL}$ ($\hmp$) is the depth of the VL sample, 
$M_{VL}$ is
the  absolute magnitude limit of the VL sample, 
$N$ the number of galaxies
contained, $\langle n \rangle = N/V $ where 
$V=(\Omega/3)R_{VL}^3$, $R_s$  ($\hmp$) is 
the effective depth and $D$ is the estimated fractal 
dimension in the range $[\langle \Lambda \rangle,R_u]$.
\label{tabvl}}
\end{table}

\clearpage

\figcaption[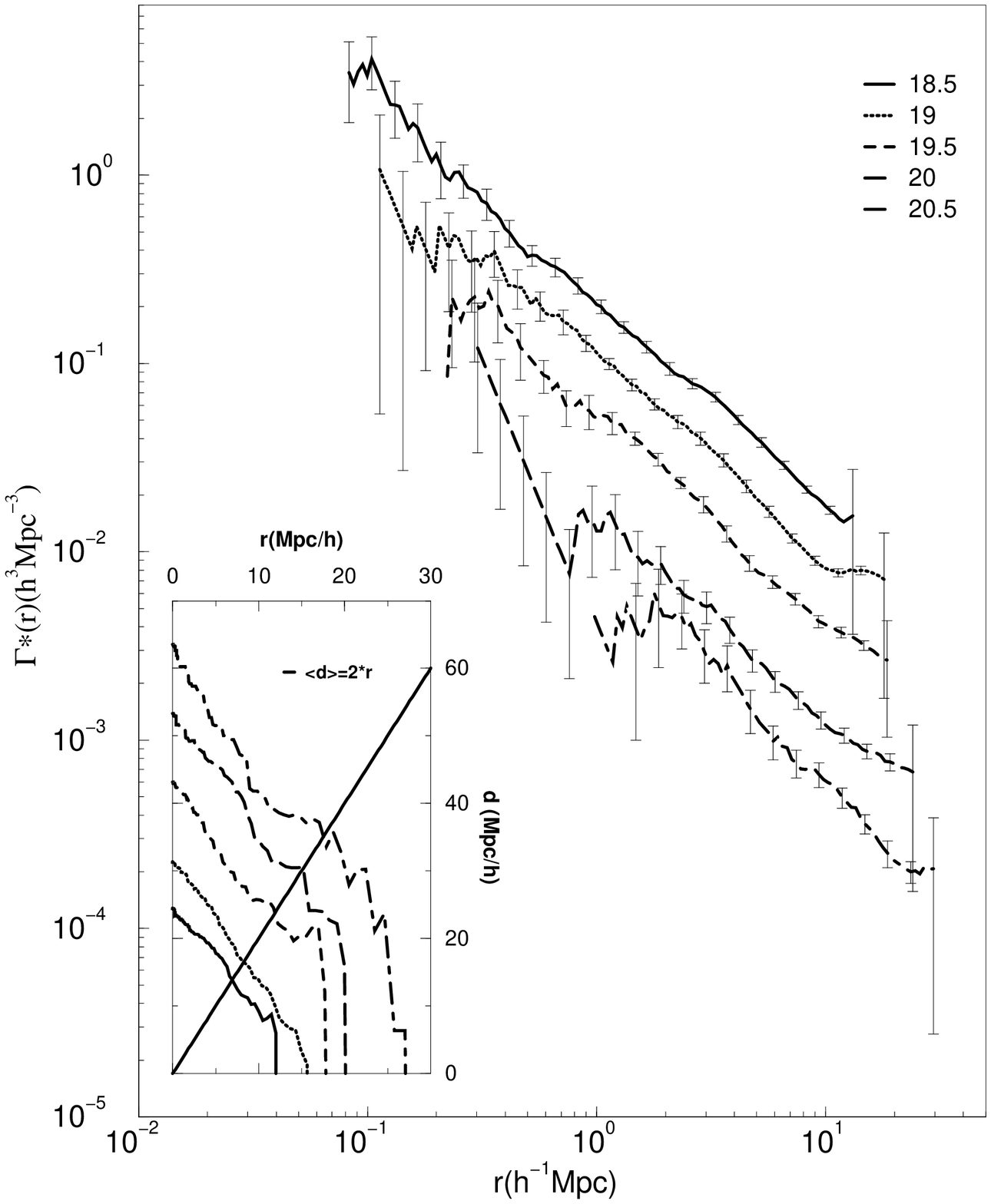]
{\label{fig1} 
The conditional average density $\Gamma^*(r)$ in the different  VL 
samples of the CfA2-South
galaxy sample. The error bars displayed correspond to the variance 
measured for 20 bootstrap 
resampling of each sample. The values of the power law fit
are reported in Tab.1. In the inserted panel we show 
the average distance between the centers of the spheres
(see text): the solid line 
corresponds to the curve $  \langle d \rangle =2r$
where $\langle d \rangle $ is the average separation between the
the centers of the spheres and $r$ is the depth.
}


\clearpage

\plotone{figure1.ps}
 

\begin{thebibliography}{}





\bibitem[Benoist et al., 1996] {ben96}
Benoist C., \etal
1996 {\it Astrophys. J.} {\bf 472}, 452-459 










\bibitem[Coleman \& Pietronero 1992]
{cp92}Coleman, P.H. and Pietronero, L.,1992
 {\it Phys.Rep.} {\bf 231},311-391




\bibitem[Davis 1997]
{dav97}Davis, M., 
 1997 p.50-60   in the Proc. of the  
Conference {\it ``Critical Dialogues in Cosmology'' }N. Turok Ed.  
World Scientific 


\bibitem[Davis \& Peebles, 1983]
{dp83} Davis, M., Peebles, P. J. E.  
1983 {\it Astrophys. J.}, {\bf 267},  465-482


\bibitem[Davis et al., 1988]
{dav88}Davis M. \etal,
 1988 {\it Astrophys. J. Lett.}  {\bf 333}, L9-L12

 


\bibitem[Huchra et al., 1998]
{huchra98} Huchra J.P., Vogeley M.S. and Geller M.J. 
1998 To Appear in {\it Astrophys.J. Suppl.} 



\bibitem[Kaiser, 1984]
{kai84} Kaiser N., 
1984 {\it Astrophys.J.} 
{\bf 284}, L9-L12




\bibitem[Mandelbrot, 1977]
{man77} Mandelbrot, B.B., 1977 
{\it ``Fractals:Form, Chance and Dimension''}, W.H.Freedman






\bibitem[Park et al., 1994]
{par94}Park C., Vogeley M., Geller M. and Huchra J.,  
1994
{\it  Astrophys. J.} {\bf 431}, 569-585 


\bibitem[Peebles, 1980]
{pee80} 
Peebles, P.E.J., 1980 {\it  ``The Large Scale Structure of 
The Universe'' }
 Princeton Univ.Press. 

\bibitem[Peebles, 1993]
{pee93}Peebles P. J. E., 1993
{\it ``Principles of physical cosmology''}, Princeton Univ. Press

\bibitem[Pietronero 1987]
{pie87} Pietronero L., 
1987  {\it   Physica A}, {\bf 144}, 257-284

\bibitem[Pietronero et al., 1997]
{pmsl97} Pietronero, L., Montuori, M. and  
Sylos Labini, F., 
1997 p.24-49 
in the Proc of the  
Conference {\it ``Critical Dialogues in Cosmology''} N. Turok Ed.  
World Scientific 

\bibitem[Sylos Labini et al., 1998]
{slmp98} Sylos Labini F., Montuori M., 
 Pietronero L., 
1998 {\it  Phys.Rep.} {\bf 293}, 66-226 

\bibitem[Sylos Labini \& Montuori, 1998]
{slm98} Sylos Labini F. and  Montuori M., 
1998 
{\it Astronom.Astrophys.}
{\bf 331}, 809-814




\bibitem[Wu et al., 1998]
{wu98}Wu K.K.S., Lahav O. and Rees M., 
1998  {\it Nature  } submitted
(astro-ph/9804062)

 

 

\end{thebibliography}
\end{document}